\documentclass[twocolumn]{book}
\usepackage[dvips]{graphicx,color}
\usepackage{makeidx,universe}
\usepackage{amsmath} 

\makeauthorindex

\BookTitle{Proceedings of the XXIX PHYSICS IN COLLISION}

\CopyRight{\copyright 2009 by Universal Academy Press, Inc.}

\RequirePackage{xspace}
\def\Dbar    {\kern 0.2em\overline{\kern -0.2em D}{}\xspace}

\def\Dz      {\ensuremath{D^0}\xspace}
\def\Dzb     {\ensuremath{\Dbar^0}\xspace}
\def\DzDzb   {\ensuremath{\Dz {\kern -0.16em \Dzb}}\xspace}

\def\fr{\frac}

\begin{document} 

\pagenumbering{arabic}

\chapter{%
{\LARGE \sf
Review on \Dz-\Dzb mixing }  \\
{\small \it To appear in the proceedings for XXIX Physics in Collision, International Symposium,\\ Kobe, Japan, August 30 - September 2, 2009}\\
{\normalsize \bf 
Fernando Mart\'{\i}nez-Vidal, }{\normalsize on behalf of the BaBar Collaboration } \\
{\small \it \vspace{-.5\baselineskip}
Instituto de F\'{\i}sica Corpuscular (IFIC), Universitat de Val\`encia-CSIC,\\
Apartado de Correos 22085, E-46071 Valencia, Spain 
}
}


\AuthorContents{F. Mart\'{\i}nez-Vidal}

\AuthorIndex{Mart\'{\i}nez-Vidal}{F.}

  \baselineskip=10pt 
  \parindent=10pt    

\section*{Abstract} 

We review the current status of \Dz-\Dzb mixing, with special emphasis in the most recent results. 
We begin with a discussion
of charm mixing and CP violation phenomenology, the evolution with the decay proper time, and physics processes contributing to these.
Then we follow with the summary of the main experimental techniques and the results in the various final states. 
We make use of the analysis reporting the first evidence of \Dz-\Dzb mixing
by BaBar in $\Dz \to K^+ \pi^-$ decays as a textbook example, and then we discuss the results from other two-body and three-body final
states. We conclude with the combination of all experimental results. Time-integrated CP violation measurements are not discussed here.

\section{Introduction} 

Particle-antiparticle oscillation (also referred as mixing) is a well known phenomenon observed in the kaon system in 1956~\cite{ref:K0mixing}, 
in the $B^0_d$ system in 1987~\cite{ref:B0dmixing}, and more recently in 2006 in the $B_s^0$ system~\cite{ref:B0smixing}. 
Mixing and CP violation (CPV) in the charm sector were first discussed over three decades ago~\cite{ref:Pais1975}, 
but experimental evidence for oscillation has been presented only in the last two 
years~\cite{ref:Kpi-BaBar,ref:KKpipi-Belle-2007,ref:Kpi-CDF,ref:KKpipi-BaBar-2008,ref:KKpipi-BaBar-2009,ref:Kpipi0-BaBar},
and no evidence for CPV has yet been reported, with upper limits currently at about 1\% level.

Charm mixing is the only involving down-type quarks in the mixing loop, since
neutral pions do not oscillate and the top quark does not have bound states. Thus \Dz-\Dzb mixing offers an unique probe 
for New Physics (NP) via flavor changing neutral currents (FCNC) in the down-quark sector, 
providing interesting constraints on NP models. The caveat is how to distinguish NP from 
Standard Model (SM) long-distance (non-perturbative) uncertainties. A possible avenue is correlating charm mixing studies 
(and possibly also rare charm decays) with a comprehensive account of CP violation in \Dz-\Dzb mixing (both within the SM and beyond).
The two effects are heavily suppressed in the SM (charm mixing is about two orders of magnitude slower than in the neutral-kaon system and CPV
is well below the per mille level), which makes these experimentally difficult to observe, although NP can produce significant enhancements.

\section{Charm mixing phenomenology} 

Neutral-$D$ mesons are created as flavor eigenstates of strong interactions, but they mix through weak interactions.
The time evolution is obtained by solving the time-dependent Schr$\ddot{\rm o}$dinger equation,

\begin{eqnarray}
i \frac{\partial}{\partial t}  \begin{bmatrix}\Dz(t)\\\Dzb(t)\\\end{bmatrix}
& = & 
{\mathbf H_w}  \begin{bmatrix}\Dz(t)\\\Dzb(t)\\\end{bmatrix},
\end{eqnarray}
with ${\mathbf H_w} = {\mathbf M} - i{\mathbf \Gamma}/2$ the effective Hamiltonian,
where ${\mathbf M}$ and ${\mathbf \Gamma}$ are $2\times2$ matrices that represent transitions via off-shell (dispersive) and on-shell
(absorptive) intermediate states, respectively. 
Assuming CPT invariance, we have $M_{11} = M_{22}$ and $\Gamma_{11} = \Gamma_{22}$. 
Since these matrices are Hermitian, $M_{12} = M_{21}^*$ and $\Gamma_{12} = \Gamma_{21}^*$. 
If CP in mixing is conserved, $M_{12} = M_{21}$ and $\Gamma_{12} = \Gamma_{21}$.

The physical (mass) eigenstates are linear combinations of the interaction eigenstates, 
$|D_{1,2} \rangle = p |\Dz\rangle \pm q |\Dzb\rangle$~\cite{ref:phase-convention},
with time evolution $|D_{1,2}(t) \rangle = e^{-i\lambda_{1,2} t} |D_{1,2} \rangle $, where $\lambda_{1,2} = m_{1,2} - i\Gamma_{1,2}/2$
are the eigenvalues. Here, $m_{1,2}$ ($\Gamma_{1,2}$) represent the mass (decay width) of the physical states. The complex
mixing parameters $p$ and $q$ obey the normalization condition $|p|^2+|q|^2=1$, and their ratio is
\begin{eqnarray}
  \fr qp & = & \pm \sqrt{\frac{M_{12}^*-i\Gamma_{12}^*/2}{M_{12}-i\Gamma_{12}/2}} = \arrowvert \fr qp \arrowvert e^{-i \phi},
\end{eqnarray}
where $\phi$ is the CP-violating phase in \Dz-\Dzb mixing. 

The time-dependent amplitude for a \Dz or \Dzb decaying into a final state $f$ after a time $t$ is
\begin{eqnarray}
\langle f | H | \begin{matrix}\Dz\\\Dzb\\\end{matrix} (t) \rangle & = & 
\fr12 \left\{ A_f g_\pm(t) + \fr qp \overline{A}_f g_\mp(t) \right\},
\end{eqnarray}
where $A_f = \langle f | H | \Dz \rangle$ and $\overline{A}_f = \langle f | H | \Dzb \rangle$ are the decay amplitudes at $t=0$, and
$g_\pm(t) = e^{-i\lambda_1 t} \pm e^{-i\lambda_2 t}$. The corresponding time evolution probability is~\cite{ref:formalism,ref:formalism2}
\begin{eqnarray}
\label{eq:master}
\Gamma\left( \begin{matrix}\Dz\\\Dzb\\\end{matrix} \to f \right) (t) = 
  e^{- \Gamma t} |A_f|^2 
\left[ 
           \begin{matrix}C_y\\\overline{C}_y\\\end{matrix} \cosh(y\Gamma t) + \right. ~~~~~~ \nonumber \\
 \left. \begin{matrix}C_x\\\overline{C}_x\\\end{matrix} \cos(x\Gamma t) +
           \begin{matrix}S_y\\\overline{S}_y\\\end{matrix} \sinh(y\Gamma t) +
           \begin{matrix}S_x\\\overline{S}_x\\\end{matrix} \sin(x\Gamma t)
\right],
\end{eqnarray}
where
\begin{eqnarray}
C_y = \frac{1+|\lambda_f|}{2} & , & C_x = \frac{1-|\lambda_f|}{2},\nonumber\\
S_y =-\Re\lambda_f & , & S_x=\Im\lambda_f,\nonumber\\
\left( \overline{C}_y, \overline{S}_y, \overline{C}_x, \overline{S}_x, \right) & = & \arrowvert \fr pq \arrowvert^2 \left( C_y,S_y,-C_x,-S_x \right),
\end{eqnarray}
with the definitions
\begin{eqnarray}
\label{eq:defPars}
  x = \frac{m_1 - m_2}{\Gamma},~~y = \frac{\Gamma_1 - \Gamma_2}{2\Gamma},~~\Gamma = \frac{\Gamma_1 + \Gamma_2}{2}, \nonumber\\
  \lambda_f = \fr qp \frac{\overline{A}_f}{A_f} = \arrowvert \fr qp \arrowvert r_d e^{-i(\Delta_f+\phi)}.~~~~~~~~~ 
\end{eqnarray}
Here $\Delta_f$ is the relative phase between $\overline{A}_f$ and $A_f$, 
and $r_d$ is the magnitude of the ratio between the two amplitudes.  
Mixing will occur either if $x$ or $y$ is non zero, while CP violation in mixing is signaled by $p \ne q$, which
can occur either if $|q/p| \ne 1$ (CP violation in mixing)  or $\phi \ne 0$ (CP violation in the interference
between mixing and decay). Direct CP violation is signaled by $A_f \ne \overline{A}_{\bar f}$.

\begin{figure}[htb!]
\begin{center}
\begin{tabular}{cc}
\includegraphics[width=0.225\textwidth]{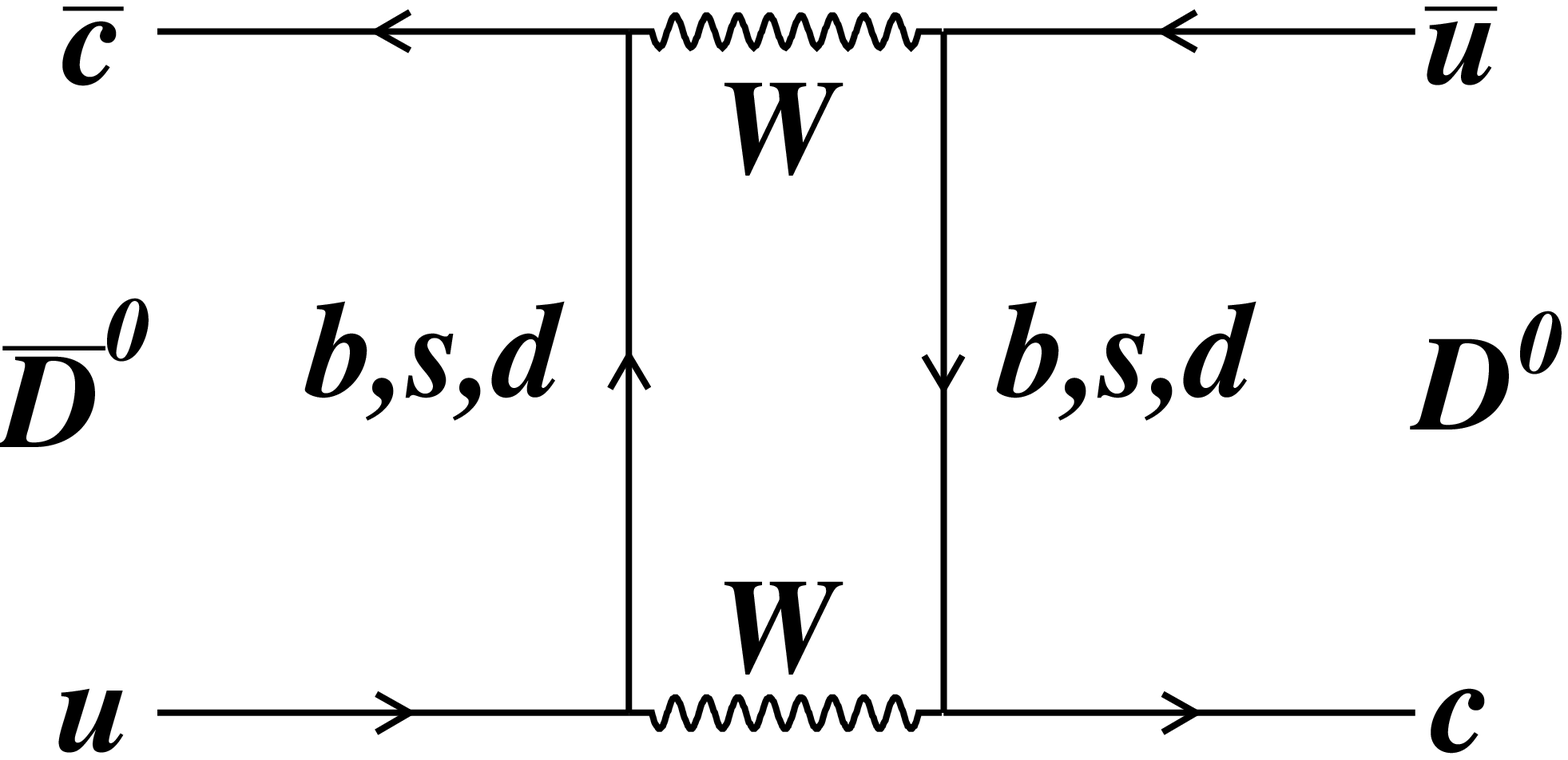} &
\includegraphics[width=0.225\textwidth]{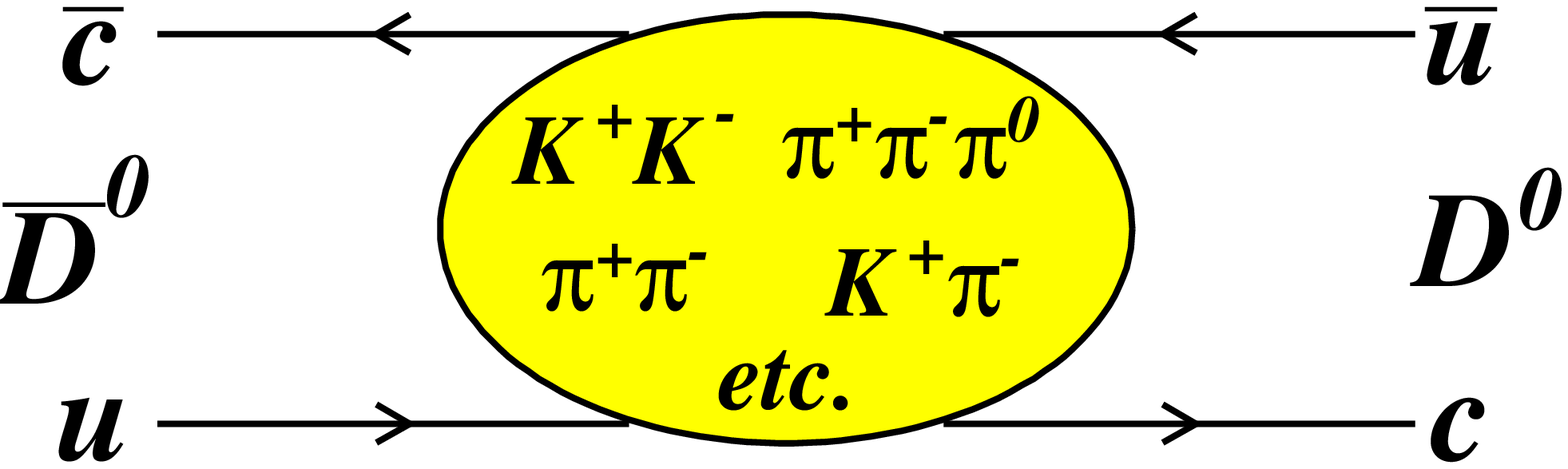} \\
\end{tabular}
\caption{\label{fig:SMprocesses} SM processes contributing to \Dz-\Dzb mixing: (Left) Short-range box diagram and 
(Right) long-range interactions with intermediate states.}
\end{center}
\end{figure}

In the SM, \Dz-\Dzb mixing arises from $|\Delta C|=2$ ($C$ is the charm quantum number) short-range box diagrams (see Fig.~\ref{fig:SMprocesses}Left)
containing down-type quarks, strongly suppressed
either by small $b$-quark couplings (CKM suppressed) or by the GIM cancellation mechanism~\cite{ref:GIM} for the light $d$- and $s$-quarks.
As a consequence, non-zero values for $x$ and $y$ are generated in the SM only at second order in ${\rm SU(3)_F}$ breaking,
$x,y \sim \sin^2\theta_C \times (m_s^2 - m_d^2)/m_c^2$, where $\theta_C$ is the Cabibbo angle and $m_s$, $m_d$, and $m_c$ are quark masses~\cite{ref:Falk2002}. 
Lowest order calculations yield $x \sim {\cal O}(10^{-5})$ and $y \sim {\cal O}(10^{-7})$, although enhancements due to higher orders in operator
product expansion (OPE) up to ${\cal O}(10^{-3})$ have been calculated~\cite{ref:DMixTheory-OPE}.
Models involving NP can greatly increase estimates for both $x$ and $y$~\cite{ref:DMixNP},
but so can $|\Delta C|=1$ long-range SM processes with intermediate states accessible 
to \Dz\ and \Dzb~\cite{ref:DMixTheory-IntermediateStates} (see Fig.~\ref{fig:SMprocesses}Right).
While most studies find $|x|,|y| < 10^{-3}$, some estimates for $x$ and $y$ allow for values as large as ${\cal O}(10^{-2})$ 
and suggest they are of opposite sign~\cite{ref:Falk2004}.
Overall, theoretical predictions for $x$ and $y$ within the SM span several orders of magnitude, reflecting
the fact that these processes are difficult to calculate~\cite{ref:formalism2,ref:DMixTheory-burdman}.

However, it would be a sign of NP if $x$ were to be significantly larger than $y$ 
or if CPV either in mixing ($p\ne q$) or decay were observed~\cite{ref:DMixNP} with current data samples.
The observation of (large) CP violation as an unambiguous sign for NP is due to the fact that all quarks building up the hadronic
states in weak decays of charm mesons belong to the first two generations. Since the $2\times 2$ Cabibbo quark-mixing matrix is real,
no CP violation is possible at tree level, and only penguin or box diagrams induced by virtual $b$-quarks can generate CP-violating
amplitudes. However, as stated above, their contributions are strongly CKM suppressed.

\section{Experimental methods} 

A generic \Dz-\Dzb mixing analysis is performed in three steps. First, the \Dz (or \Dzb) flavor at production time ($t=0$) is identified (``tagged'') 
using $D^{*+} \to \Dz \pi^+$ decays~\cite{ref:chargeconj}. These events are usually selected and characterized using the
invariant mass of the exclusively reconstructed \Dz meson, $m_{\Dz}$, and the mass difference between the reconstructed $D^{*+}$ and \Dz mesons, 
$\Delta m = m(D^{*+}) - m_{\Dz}$. The distribution of $\Delta m$ shows a narrow peak, due to the small $Q$-value of the $D^{*+} \to \Dz \pi^+$ decay.
Other tools used to improve the event selection and reduce backgrounds are particle identification (for leptons, kaons or pions, depending
on the \Dz final state) and cuts on \Dz (high) and soft pion (low) momentum.
The charge of the soft pion from the $D^*$ decay unambiguously identifies the \Dz flavor at production.
Then, the \Dz flavor at decay time is identified using the charge of the final state particles. 
For example, if the reconstructed final state is a positive kaon and a negative pion, 
and the soft pion from the $D^*$ decay has a negative charge, 
then we have a ``Right sign (RS)'' combination, $\Dzb \to K^+ \pi^-$. On the contrary, if the $D$ meson has been tagged as \Dz, then we have a
``Wrong sign (WS)'' combination, $\Dz \to K^+ \pi^-$. When we have a tagged \Dz meson at production, a WS combination can occur either if
the \Dz meson decays via a double-Cabibbo-suppressed (DCS) transition or if it oscillates into a \Dzb meson followed by decay through a Cabibbo favored (CF)
transition. Only a time-dependent analysis of the WS rate allows to distinguish between these two effects.
Finally, the production and decay vertices of the \Dz meson are reconstructed in order 
to calculate the decay flight length and hence the decay proper time $t$ and its uncertainty $\sigma_t$. 
At B factories, restricting the production point to the luminous region of the collider (beam spot) greatly improves the precision on the 
decay time reconstruction, as well as on $\Delta m$.
At these facilities the average decay length is about 240 $\mu$m, with typical 
resolution about 100 $\mu$m (the latter depends on the specific reconstructed final state). 
Analyses usually apply quality cuts
on the proper-time error, in order to reduce effects from wrongly reconstructed vertices.

The tagging can also be performed using coherent \Dz-\Dzb production at charm factories running slightly above 
the \Dz-\Dzb threshold, although in this case only time-integrated mixing related measurements
are possible at present facilities~\cite{ref:formalism-coherent}. 
%
On the other hand, some analyses can be performed using tagged or untagged samples, as it is the case 
of the lifetime differences between decays to CP eigenstates (like $\Dz \to K^+ K^-$) and to the CP-mixed state $\Dz \to K^-\pi^+$,
as discussed later.

Four experimental techniques have been used to measure \Dz-\Dzb mixing, depending on the specific \Dz final state:
WS semileptonic decays, WS hadronic decays, decays to CP eigenstates, and self-conjugate three-body final states
containing a combination of quasi-two body flavor and CP eigenstates, particularly $K_S^0 \pi^+ \pi^-$. 
Quantum-correlated final states at charm factories are also sensitive to \Dz-\Dzb mixing (mostly $y$) via 
time-integrated observables~\cite{ref:formalism-coherent}, although their sensitivity is not competitive with time-dependent measurements.
These are however fundamental to provide information on magnitudes and phases of relevant amplitude ratios, as described later.

\section{Wrong sign hadronic decays} 

Sensitivity to mixing using WS hadronic decays, for example $\Dz \to K^+ \pi^-$, is obtained by analyzing their proper-time evolution.
Time-dependent studies allow separation of the direct DCS $\Dz \to K^+ \pi^-$ amplitude from the mixing contribution followed
by the CF decay, $\Dz \to \Dzb \to K^+ \pi^-$~\cite{ref:formalism,ref:formalism2}. 
Taking $|\lambda_f| << 1$ and assuming small mixing, 
\begin{eqnarray}
\label{eq:WShadronic}
\frac{\Gamma(\Dz \to f)(t)}{e^{-\Gamma t} } \propto R_D + (\Gamma t)^2 \frac{R_M}{2} + (\Gamma t)\sqrt{R_D}y',
\end{eqnarray}
where $f$ represents the WS final state, $R_D=r_d^2$ is the ratio of DCS to CF decay rates, $R_M = (x^2 + y^2)/2$ is the mixing rate,
and $y' = y \cos\delta_f - x\sin\delta_f$, where $\delta_f = - \Delta_f$ is the relative strong phase between the DCS and CF decay amplitudes.
The minus sign originates from the sign of $V_{us}$ relative to $V_{cd}$, where $V$ denotes the quark-mixing CKM matrix.
In Eq.~(\ref{eq:WShadronic}),
the first term corresponds to the DCS contribution to the WS rate (time independent), the second term is the contribution 
from mixing,
and the third term is the interference between 
mixing and CF decays. Since $x$ and $y$ are small, ${\cal O}(10^{-2})$, it is precisely the interference term (linear in decay time, $x$ and $y$)
which gives the best sensitivity to mixing through $y'$. Let us note that $R_M \equiv (x'^2 + y'^2)/2$,
with $x' = x \cos\delta_f + y\sin\delta_f$. However, a direct extraction of $x$ and $y$ from Eq.~(\ref{eq:WShadronic}) is not
possible due to the unknown relative phase $\delta_f$.

Searches for charm mixing in WS $\Dz \to K^+ \pi^-$ decays have been performed by the experiments E971 
(using $2\times10^{10}$ events from $\pi^- N$ interactions at 500 GeV) and FOCUS (from $10^6$ $D\to Kn(\pi)$ events 
from $\gamma N$ interactions), and by CLEO from 9 fb$^{-1}$ of $e^+e^- \to \Upsilon(4S)$ data~\cite{ref:Kpi-E791-FOCUS-CLEO}.
However, the available statistics from these experiments was not enough to obtain evidence of mixing.

The first evidence for \Dz-\Dzb mixing in WS $\Dz \to K^+ \pi^-$ decays has been reported by BaBar using 384 fb$^{-1}$ of data~\cite{ref:Kpi-BaBar}. 
The simultaneous fit to the RS and WS data samples to describe the signal and the random soft pion and misreconstructed \Dz background components 
yields $1141500\pm1200$ and $4030\pm90$ signal events, respectively. 
Thus the fraction of WS decays is measured to be $R_{WS}=[0.353\pm0.008{\rm (stat.)}\pm0.004{\rm (syst.)}]\%$. In the presence of mixing, 
$R_{WS} > R_D$, as can easily be obtained integrating Eq.~(\ref{eq:WShadronic}).

\begin{figure}[htb!]
\begin{center}
\includegraphics[width=0.4\textwidth]{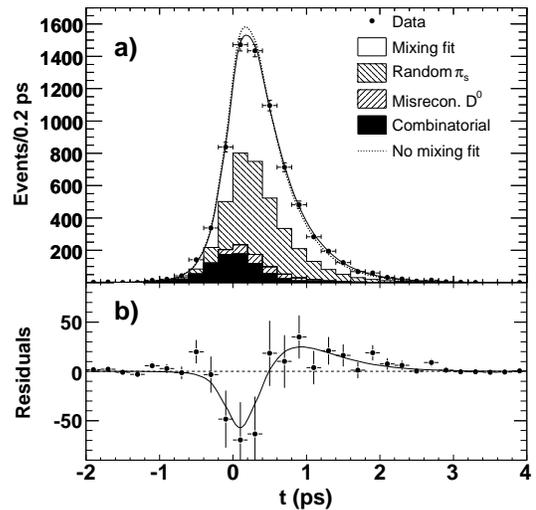}
\caption{\label{fig:Kpi-BaBar-time} BaBar WS $\Dz \to K^+ \pi^-$ analysis~\cite{ref:Kpi-BaBar}.
(a) Projections of the proper-time distribution of WS candidates and fit results allowing (solid curve) and 
not allowing (dashed curve) mixing. (b) The points represent the difference between the data and the no-mixing fit, and the curve shows the
difference between the fits with and without mixing.}
\end{center}
\end{figure}
\begin{figure}[htb!]
\begin{center}
\includegraphics[width=0.4\textwidth]{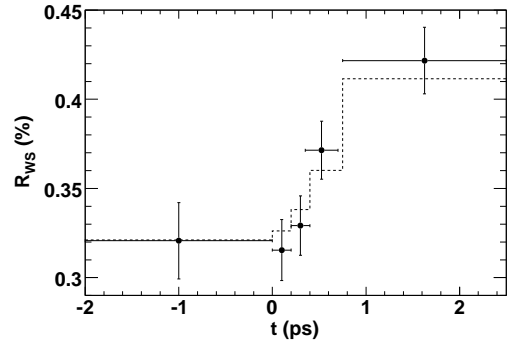}
\vskip-2.5cm
\caption{\label{fig:Kpi-BaBar-WSratio} BaBar WS $\Dz \to K^+ \pi^-$ analysis~\cite{ref:Kpi-BaBar}.
The measured WS fraction in slices of measured proper time (points). The dashed line shows the expected WS fraction
as determined from the mixing fit. In the absence of mixing, no time dependence would be observed.}
\end{center}
\end{figure}

The measured proper-time distribution for the WS signal is modeled by Eq.~(\ref{eq:WShadronic}) convolved with a resolution function determined 
using the RS proper-time fit. The proper-time distribution for WS data in the $m_{\Dz}-\Delta m$ signal box is shown in Fig.~\ref{fig:Kpi-BaBar-time}, 
together with the fit results with and without mixing, shown as the overlaid curves. 
The mixing parameters are 
$y' = [0.97\pm0.44{\rm (stat.)}\pm0.31{\rm (syst.)}]\%$ 
and 
$x'^2 = [-0.022\pm0.030{\rm (stat.)}\pm0.021{\rm (syst.)}]\%$,
and a correlation between them of $-0.95$. The ratio of DCS to CF decay
rates is measured to be 
$[0.303\pm0.016{\rm (stat.)}\pm0.010{\rm (syst.)}]\%$.
The systematic uncertainties are dominated by the signal resolution function as extracted from the RS sample. 
As expected, $R_{WS} > R_D$, revealing the presence of mixing. As another cross-check of the mixing signal, $R_{WS}$ can also be measured in
slices of proper time, repeating the fit to the RS and WS data samples in each of these slices.
The fitted and expected WS fractions are shown in Fig.~\ref{fig:Kpi-BaBar-WSratio} and are seen to increase quadratically
with time, as expected according to Eq.~(\ref{eq:WShadronic}). 
The significance of the mixing signal is equivalent to $3.9\sigma$ or $9.6\times10^{-5}$ confidence level (CL), where $\sigma$ denotes one standard deviation. 
Separate proper-time fits to \Dz and \Dzb events allow to determine a CP-violating asymmetry 
$A_D = (R_D^+ - R_D^-)/(R_D^+ + R_D^-) = [-2.1\pm5.2{\rm (stat.)}\pm1.5{\rm (syst.)}]\%$, where
$R_D^+(R_D^-)$ is the ratio of DCS and CF decay rates for $\Dz(\Dzb)$, thus no evidence for CP violation is observed.

\begin{figure}[htb!]
\begin{center}
\includegraphics[width=0.45\textwidth]{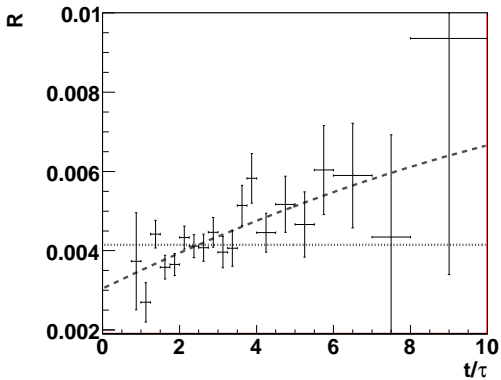}
\vskip-2.0cm
\caption{\label{fig:Kpi-CDF-WSratio} CDF WS $\Dz \to K^+ \pi^-$ analysis~\cite{ref:Kpi-CDF}.
The measured ratio of WS to RS decays as a function of the decay proper time (points). The dashed curve is the result
of the mixing fit, while the dotted line is a fit assuming no mixing.
}
\end{center}
\end{figure}

These results have been confirmed by the CDF experiment using a data sample of 1.5 fb$^{-1}$ of $p\bar{p}$ interactions 
at $\sqrt{s}=1.96$~TeV~\cite{ref:Kpi-CDF}. The analysis is similar to that from BaBar, although the different production environment 
makes the details to differ significantly.
The time-integrated fit to the RS and WS data samples yield $(3.044\pm0.002)\times10^6$ and $(12.7\pm0.3)\times10^3$ signal events, 
respectively. The ratio of WS to RS decays as a function of the decay proper time in the range between $0.75$ and $10$ \Dz lifetimes
shows again an approximately linear 
dependence, as observed in Fig.~\ref{fig:Kpi-CDF-WSratio} 
The parabolic fit of the data in this figure returns 
$y' = (0.85\pm0.76)\%$, 
$x'^2 = (-0.012\pm0.035)\%$,
and
$R_D = (0.304\pm0.055)\%$,
where the errors include statistical and systematic uncertainties.
The significance of the mixing signal is equivalent to $3.8\sigma$ ($1.5\times10^{-4}$ CL). 
These results are essentially identical to those obtained by BaBar,
in spite of the very different production environment and sources of systematic uncertainties.

An earlier search by Belle for mixing in this decay mode using 400 fb$^{-1}$ of data did not yield clear evidence for mixing~\cite{ref:Kpi-Belle}.
The time-integrated fit to the RS and WS data samples returns very similar yields as those obtained by BaBar,
$1073993\pm1108$ and $4024\pm88$ signal events, respectively, from which
$R_{WS}=[0.377\pm0.008{\rm (stat.)}\pm0.005{\rm (syst.)}]\%$. The time-dependent mixing fit 
yields
$y' = (0.06^{0.40}_{0.39})\%$, 
$x'^2 = (0.018^{0.021}_{0.023})\%$,
and
$R_D = (0.364\pm0.017)\%$,
where the errors include statistical and systematic uncertainties.
The correlation between $y'$ and $x'^2$ is $-0.91$.
The no-mixing hypothesis is excluded at $2.1\sigma$ ($3.9\%$ CL).
This result agrees with those obtained by BaBar and CDF at $2\sigma$ level.
Separate proper-time fits to \Dz and \Dzb events show no evidence for CP violation.



Quantum-correlated \Dz-\Dzb pairs produced in $\Psi(3770)$ decays at charm factories, with definite charge-conjugation eigenvalue $C=-1$,
can be exploited to make a determination of the relative strong phase $\delta_{K\pi}$ to translate the 
measurement of $y'$ into $y$~\cite{ref:CLEOc}. At slightly higher energies (above $DD^*$ threshold) one can also produce such 
pairs with $C=+1$ (additional photons in the final state).
One can use the fact that heavy-meson pairs produced in the decays of heavy-quarkonium states have the property that 
the two mesons are in CP- or flavor-correlated states~\cite{ref:formalism-coherent}. 
For instance, one may tag one of the neutral-$D$ mesons as a CP eigenstate through its decay into CP eigenstates,
such $K_S(\pi^0,\rho^0,\omega,\eta,\eta',\phi)$, $K^+ K^-$, and $\pi^+ \pi^-$. The other neutral-$D$ meson
must then have opposite CP if $C(\Dz\Dzb)=-1$ and the same CP if $C(\Dz\Dzb)=+1$. Then one measures its decay
rate into $K^+ \pi^-$, which includes, as discussed, an interference between CF and DCS amplitudes. The measured rate
thus depends on the CF and DCS rates and the relative strong phase $\delta_{K\pi}$.
More generally, one can measure time-integrated yields of correlated (``double tags'') and uncorrelated (``single tags'') neutral-$D$ meson decays
to CP eigenstates (CP-even and CP-odd) and flavor eigenstates (semileptonic and hadronic decays, with leptons and/or kaons as final state particles).
The ratio of correlated and uncorrelated decay rates depends on mixing parameters $R_D$, $\sqrt{R_D} \cos \delta_f$, $y$, $x^2$, and 
$\sqrt{R_D} x \sin \delta_f$. From 818 pb$^{-1}$ of data recorded at $\Psi(3770)$ (and at slightly higher energies) and using
branching ratios from other experiments, CLEOc obtains
$\cos \delta_{K\pi} = 1.03^{+0.31}_{-0.17}{\rm (stat.)}\pm0.06{\rm (syst.)}$~\cite{ref:CLEOc}.
The extraction of other mixing parameters is not competitive with time-dependent methods.

Further evidence for \Dz-\Dzb mixing has been reported by BaBar using a time-dependent Dalitz plot analysis of 
the multi-body WS decay $\Dz \to K^+ \pi^- \pi^0$~\cite{ref:Kpipi0-BaBar}. The analysis in such decays is formally
similar to the WS $\Dz \to K^+ \pi^-$, but now the decay rate is a function of both the decay proper time and 
the Dalitz plot variables $s_0 = m^2_{K^+\pi^-}$ and $s_+ = m^2_{K^+\pi^0}$,
\begin{eqnarray}
\label{eq:WShadronicKpipi0}
\frac{\Gamma(\Dz \to f)(s_0,s_+,t)}{e^{-\Gamma t} } = ~~~~~~~~~~~~~~~~~~~~~~~~~~~ \nonumber \\
   |A_f|^2 + |\overline{A}_f|^2 (\Gamma t)^2 \frac{R_M}{2} + (\Gamma t) |A_f| |\overline{A}_f| y',
\end{eqnarray}
where $A_f(s_0,s_+)$ is the DCS amplitude, 
$\overline{A}_f(s_0,s_+)$ is the CF amplitude,
and $y' = y \cos \delta_f(s_0,s_+) - x \sin \delta_f(s_0,s_+)$, 
with $\delta_f(s_0,s_+) = {\rm arg} [ A_f^*(s_0,s_+) \overline{A}_f(s_0,s_+) ]$ the relative strong phase 
between the DCS and CF amplitudes, now varying with the Dalitz plot position.
As it can be seen in Eq.~(\ref{eq:WShadronicKpipi0}), the sensitivity to mixing comes from the variation 
of the Dalitz plot distribution with time produced by the CF-mixing interference term, 
which in turn mainly depends on the interference between the CF $\Dzb \to K^+ \rho^-$ and DCS $\Dz \to K^{*+} \pi^-$ amplitudes, 
since these decays dominate the RS and WS Dalitz plots, respectively.

\begin{figure}[htb!]
\begin{center}
\begin{tabular}{cc}
\includegraphics[width=0.22\textwidth]{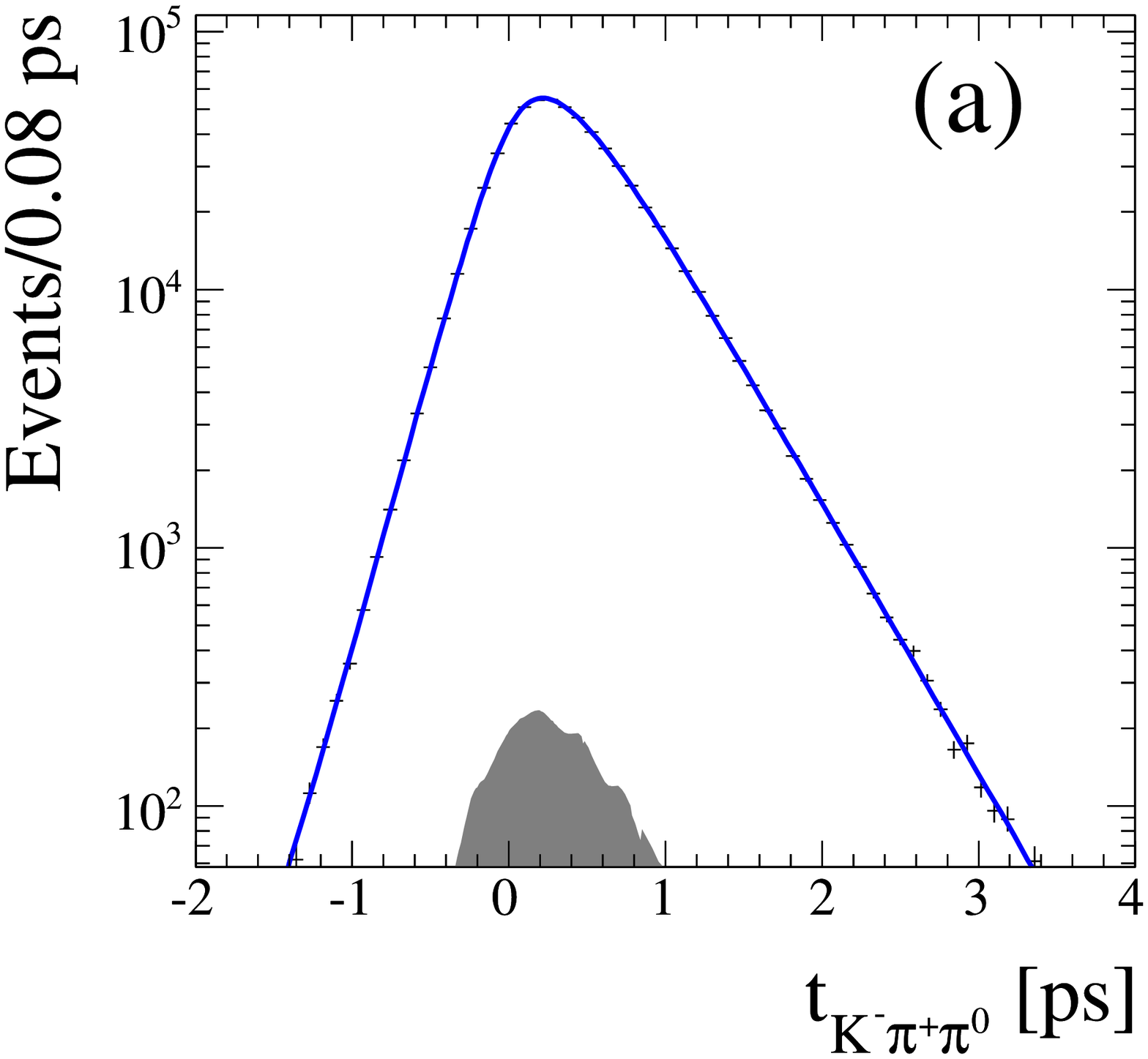} &
\includegraphics[width=0.22\textwidth]{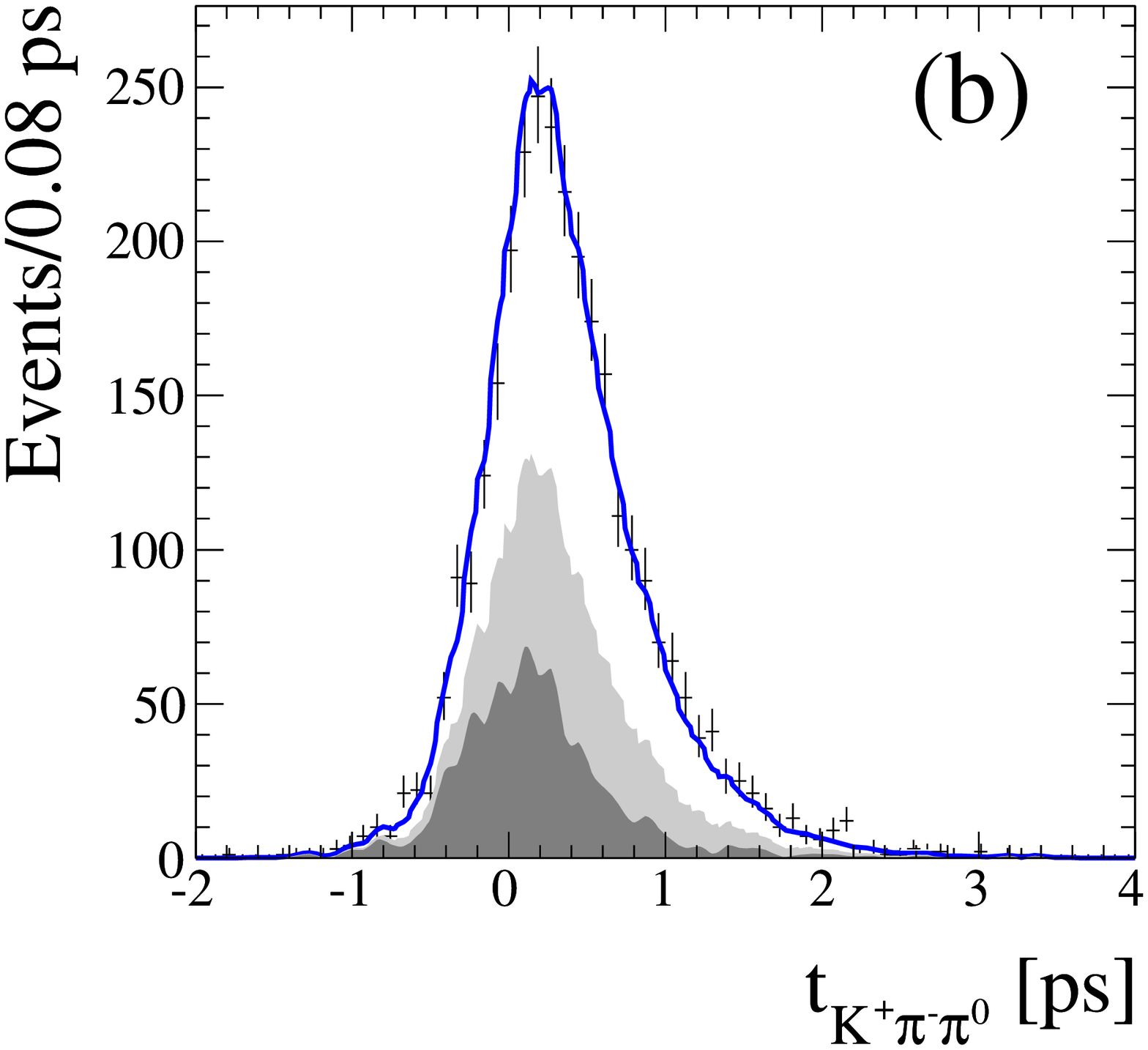} \\
\includegraphics[width=0.22\textwidth]{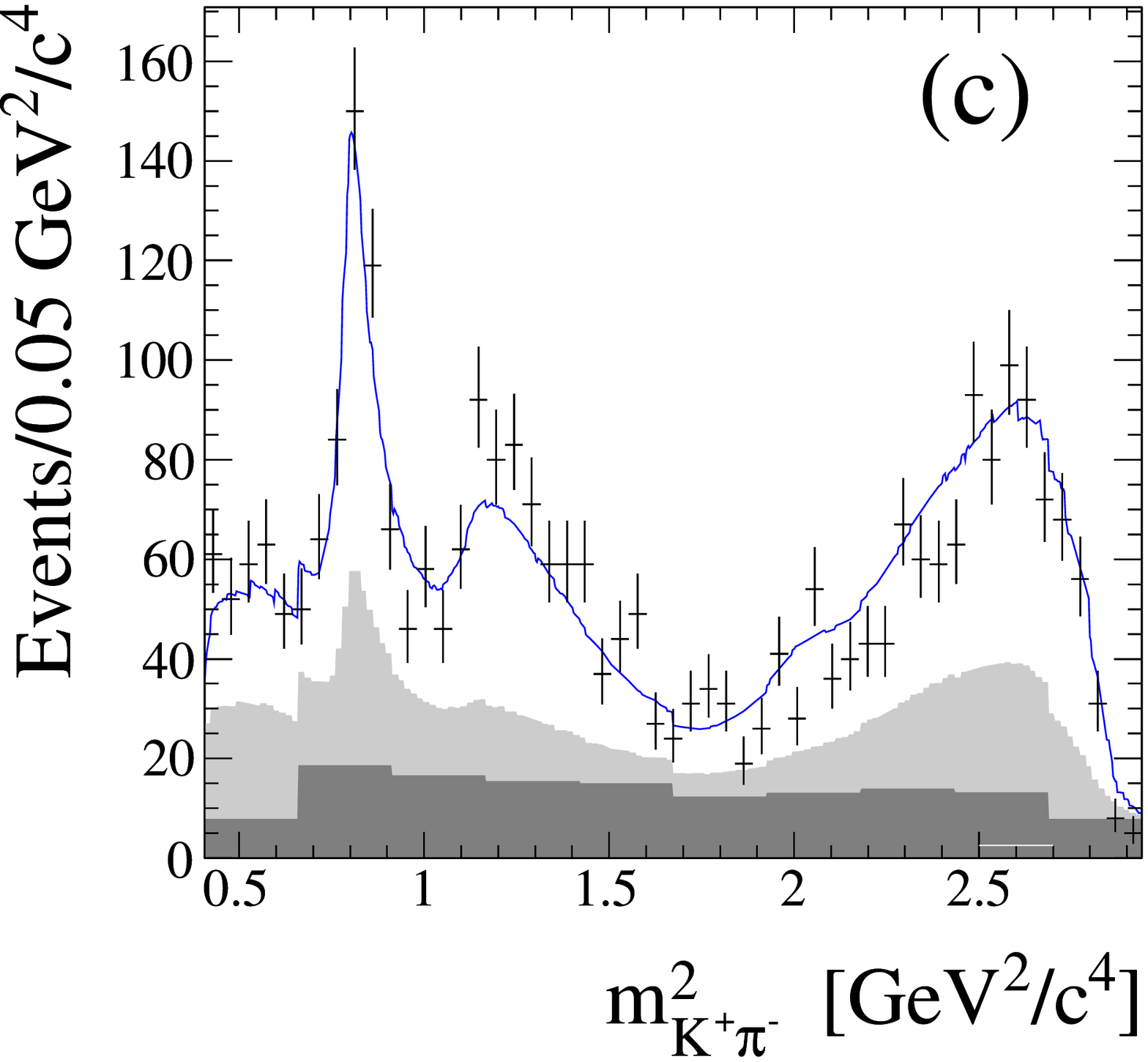} &
\includegraphics[width=0.22\textwidth]{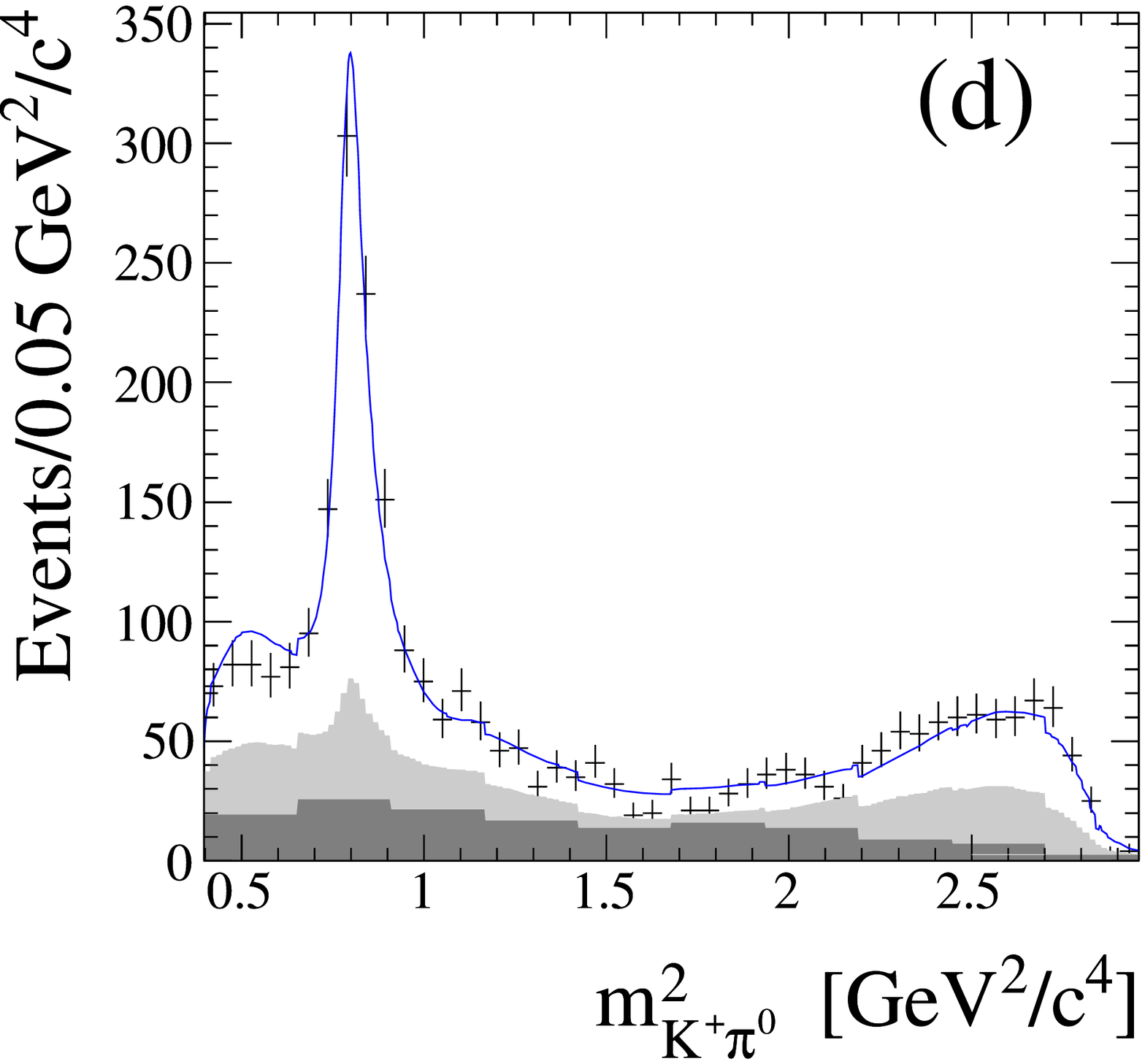} \\
\end{tabular}
\caption{\label{fig:Kpipi0-BaBar-Projdalitz} BaBar WS $\Dz \to K^+ \pi^- \pi^0$ analysis~\cite{ref:Kpipi0-BaBar}.
Proper-time distribution for (a) RS and (b) WS decays with the fit projection overlaid. Projections of the Dalitz
plot distributions for (c) $s_0$ and (d) $s_+$. The gray histograms represent the random soft pion background, while the
dark histograms show the misreconstructed \Dz background.}
\end{center}
\end{figure}

BaBar determined the CF amplitude $\overline{A}_f$ in a time-integrated Dalitz 
plot analysis of the RS decay sample, consisting of 658,986 events with a purity of 99\%. 
This amplitude is then used in the analysis of the WS sample, containing 3009 events with a purity of 50\%,
where the DCS amplitude $A_f$ is extracted along with the mixing parameters. Each of the amplitudes
$\overline{A}_f$ and $A_f$ are in turn described as a coherent sum of amplitudes, each describing a separate resonance 
(the usual isobar approach). Figure~\ref{fig:Kpipi0-BaBar-Projdalitz} shows the RS and WS proper-time distributions as well as the projections on $s_0$ and $s_+$.
Since for both $\overline{A}_f$ and $A_f$ one complex amplitude must be fixed arbitrarily and the CF and DCS 
Dalitz plots are different, the sensitivity to $x$ and $y$ is in the form
$y' = y \cos\delta_{K\pi\pi^0} - x\sin\delta_{K\pi\pi^0}$ and 
$x' = x \cos\delta_{K\pi\pi^0} + y\sin\delta_{K\pi\pi^0}$, where
$\delta_{K\pi\pi^0}$ is the strong phase difference between the DCS $\Dz \to K^+ \rho^-$ and the CF $\Dzb \to K^+ \rho^-$ amplitudes.
This phase is unknown and different from $\delta_{K\pi}$.
The measured mixing parameters are 
$x' = [2.61^{+0.57}_{-0.68}{\rm (stat.)}\pm0.39{\rm (syst.)}]\%$ 
and 
$y' = [-0.06^{+0.55}_{-0.64}{\rm (stat.)}\pm0.34{\rm (syst.)}]\%$,
and a correlation between them of $-0.75$. 
The significance of the mixing signal is equivalent to $3.2\sigma$ ($0.1\%$ CL). No evidence for CP violation is seen.

\section{Hadronic decays to CP eigenstates} 

The lifetime difference between states of different CP content, for example $\Dz \to K^+ K^-$ (CP even) compared to $\Dz \to K^- \pi^+$ (CP mixed), can 
also be used to measure \Dz-\Dzb mixing. For small mixing and taking $\lambda_f \approx 1$ 
(for CP-even decays, since $\delta_f=0,\pi$ for CP$=+1,-1$)~\cite{ref:formalism,ref:formalism2},
\begin{eqnarray}
\label{eq:CPhadronic}
\Gamma\left( \begin{matrix}\Dz\\\Dzb\\\end{matrix} \to f_{CP} \right) (t) \propto e^{- \Gamma_\pm t},
\end{eqnarray}
where $\Gamma_\pm = \Gamma (1+y'_\pm)$ is the \Dz or \Dzb effective lifetime, with $y'_\pm \approx y \cos \phi \mp x \sin \phi$. For untagged neutral-$D$ mesons,
\begin{eqnarray}
\label{eq:CPhadronic-untagged}
\Gamma\left( \Dz\ {\rm or}\ \Dzb\ \to f_{CP} \right) (t) \propto e^{- \langle \Gamma_\pm \rangle t},
\end{eqnarray}
where $\langle \Gamma_\pm \rangle = \Gamma (1+\langle y'_\pm \rangle)$ is the average \Dz and \Dzb effective lifetime.
We clearly observe that for $y \ne 0$, the lifetimes to CP eigenstates ($\Gamma_\pm$, $\langle \Gamma_\pm \rangle$) and CP mixed states ($\Gamma$) differ.
The experimentally defined observables are
\begin{eqnarray}
y_{CP} & = & \frac{\tau_{K^-\pi^+}}{\langle \tau_{h^+h^-}\rangle}-1 = \frac{\tau}{(\tau_+ + \tau_-)/2}-1 \approx y \cos\phi,\nonumber\\
A_\tau & = & \frac{\tau_+ - \tau_-}{\tau_+ + \tau_-} \approx x\sin\phi,
\end{eqnarray}
where $\tau_{K^-\pi^+} \equiv \tau = 1/\Gamma$ and $\langle \tau_{h^+h^-} \rangle = (\tau_+ + \tau_-)/2$ with $\tau_\pm = 1/\Gamma_\pm$ is the mean lifetime for 
neutral-$D$ mesons decaying into CP eigenstates. The observable $A_\tau$ is the asymmetry in their lifetimes, sometimes
replaced by $\Delta Y = \tau A_\tau / \langle \tau_\pm \rangle = (1-y_{CP})A_\tau$.
In the limit of vanishing CP violation $y_{CP} = y$, 
and $A_\tau$ (or $\Delta Y$) is zero.
Both $y_{CP}$ and $A_\tau$ (or $\Delta Y$) vanish if there is no \Dz-\Dzb mixing.
The measurement of $y_{CP}$ requires precise determinations of lifetimes using either 
tagged or untagged neutral-$D$ mesons, but $A_\tau$ can only be measured using tagged \Dz and \Dzb mesons.
The advantage of these observables is that most of the systematic uncertainties related to the signal cancel in the ratios, 
although background related systematic uncertainties do not.

Searches for \Dz-\Dzb mixing in hadronic decays to CP eigenstates have been done by the E971, FOCUS, and 
CLEO experiments~\cite{ref:KKpipi-E791-FOCUS-CLEO}. 
However, the first evidence for charm mixing in these decays has been
presented by Belle~\cite{ref:KKpipi-Belle-2007} simultaneously with the BaBar WS $\Dz \to K^+ \pi^-$~\cite{ref:Kpi-BaBar} evidence, 
and both constitute the chief analyses reporting the first evidences for \Dz-\Dzb mixing (quickly confirmed by CDF~\cite{ref:Kpi-CDF}).

\begin{figure}[htb!]
\begin{center}
\includegraphics[width=0.9\textwidth]{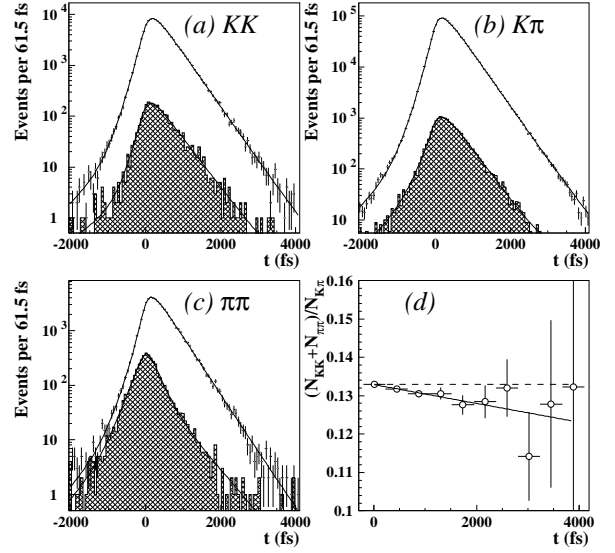}
\vskip-7.0cm
\caption{\label{fig:KK-pipi-Belle} Belle $\Dz \to K^+ K^-, \pi^+ \pi^-$ analysis~\cite{ref:KKpipi-Belle-2007}.
Decay time distributions of (a) $\Dz \to K^+ K^-$, (b) $\Dz \to K^- \pi^+$, and (c) $\Dz \to \pi^+ \pi^-$ decays.
The curves are the projections of the mixing fit, and the cross-hatched areas represent the background contribution.
(d) Ratio of decay-time distributions between $\Dz \to K^+ K^-,\pi^+ \pi^-$ and $\Dz \to K^- \pi^+$. The solid line is a linear fit to the data points.
}
\end{center}
\end{figure}

Using 540 fb$^{-1}$ of data, Belle has measured 
$y_{CP} = [1.31\pm0.32{\rm (stat.)}\pm0.25{\rm (syst.)}]\%$ employing the $K^+ K^-$ and $\pi^+ \pi^+$ final states (both CP-even), and 
found no evidence for CP violation in these decays since they obtained for the lifetime 
asymmetry $A_\tau = [0.01\pm0.30{\rm (stat.)}\pm0.15{\rm (syst.)}]\%$~\cite{ref:KKpipi-Belle-2007}. 
The proper-time distribution for the $\Dz \to K^+ K^-, K^- \pi^+, \pi^+ \pi^-$ samples, 
consisting of $111\times10^3$, $1.22\times10^6$, $49\times10^3$ signal events with purities of 98\%, 99\%, 92\%,
are shown in Fig.~\ref{fig:KK-pipi-Belle}(a,b,c), respectively, together with the projection of the fit. 
The CL of the no-mixing hypothesis ($y_{CP}=0$) is $6\times10^{-4}$, which 
corresponds to a significance of $3.2\sigma$. The mixing effect can be seen in Fig.~\ref{fig:KK-pipi-Belle}(d),
which shows the ratio of decay-time distributions for $\Dz \to K^+ K^-,\pi^+ \pi^-$ and $\Dz \to K^- \pi^+$ decays,
increasing linearly with time, as expected according to Eq.~(\ref{eq:CPhadronic-untagged}) to first order in $y$, 
$\Gamma\left( \Dz\ {\rm or}\ \Dzb\ \to f_{CP} \right) (t) \propto 1 - \Gamma y_{CP} t$.

\begin{figure}[htb!]
\begin{center}
\includegraphics[width=0.25\textwidth]{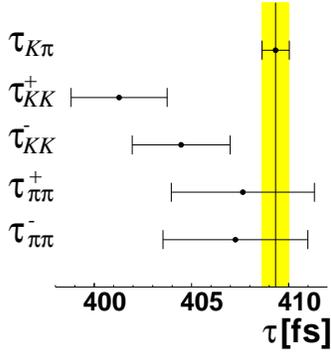} 
\vskip-1.2cm
\caption{\label{fig:KK-pipi-BaBar-Tagged} BaBar $\Dz \to K^+ K^-, \pi^+ \pi^-$ analysis~\cite{ref:KKpipi-BaBar-2008}.
Summary of the measured lifetimes in the 5 $D^{*}$ tagged $K^+ K^-$, $\pi^+ \pi^-$ and $K^- \pi^+$ samples.
}
\end{center}
\end{figure}

A very similar study has also been performed by BaBar using 384 fb$^{-1}$ of data, yielding
$y_{CP} = [1.24\pm0.39{\rm (stat.)}\pm0.13{\rm (syst.)}]\%$ and 
$\Delta Y = [-0.26\pm0.36{\rm (stat.)}\pm0.08{\rm (syst.)}]\%$~\cite{ref:KKpipi-BaBar-2008}.
These results are obtained fitting simultaneously the proper-time distributions for the $D^{*}$ tagged $K^+ K^-$, 
$\pi^+ \pi^-$, and $K^- \pi^+$ samples, consisting of $69,696$, $30,679$ and $730,880$ signal events with purities of 99.6\%, 98.0\%, 99.9\%.
The significance of the no-mixing hypothesis is of $3.0\sigma$, reflecting the significance of the difference of lifetimes
between $K^+ K^-$, $\pi^+ \pi^-$ and $K^- \pi^+$, as summarized in Fig.~\ref{fig:KK-pipi-BaBar-Tagged}

\begin{figure}[htb!]
\begin{center}
\begin{tabular}{cc}
\includegraphics[width=0.24\textwidth]{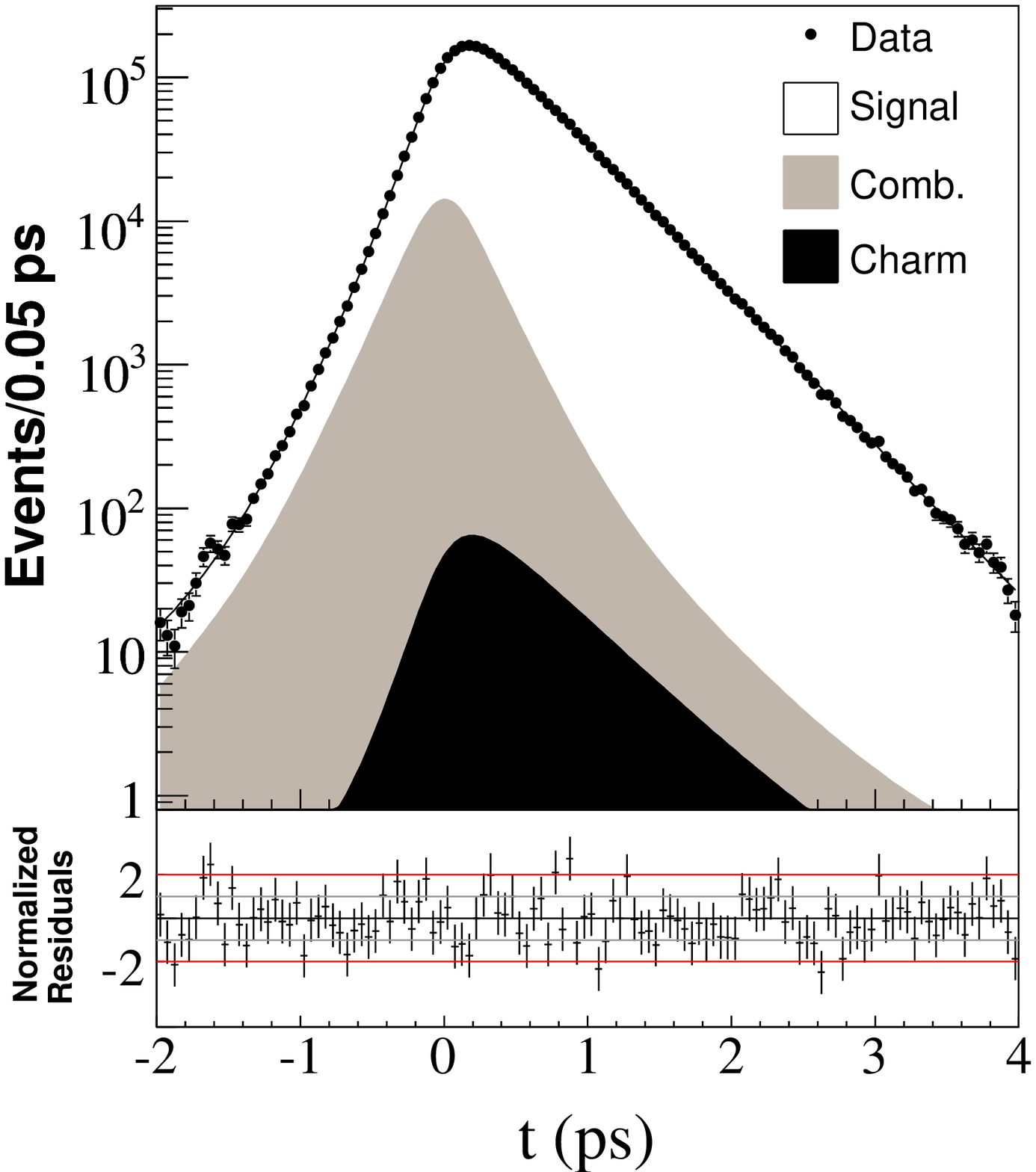} &
\includegraphics[width=0.24\textwidth]{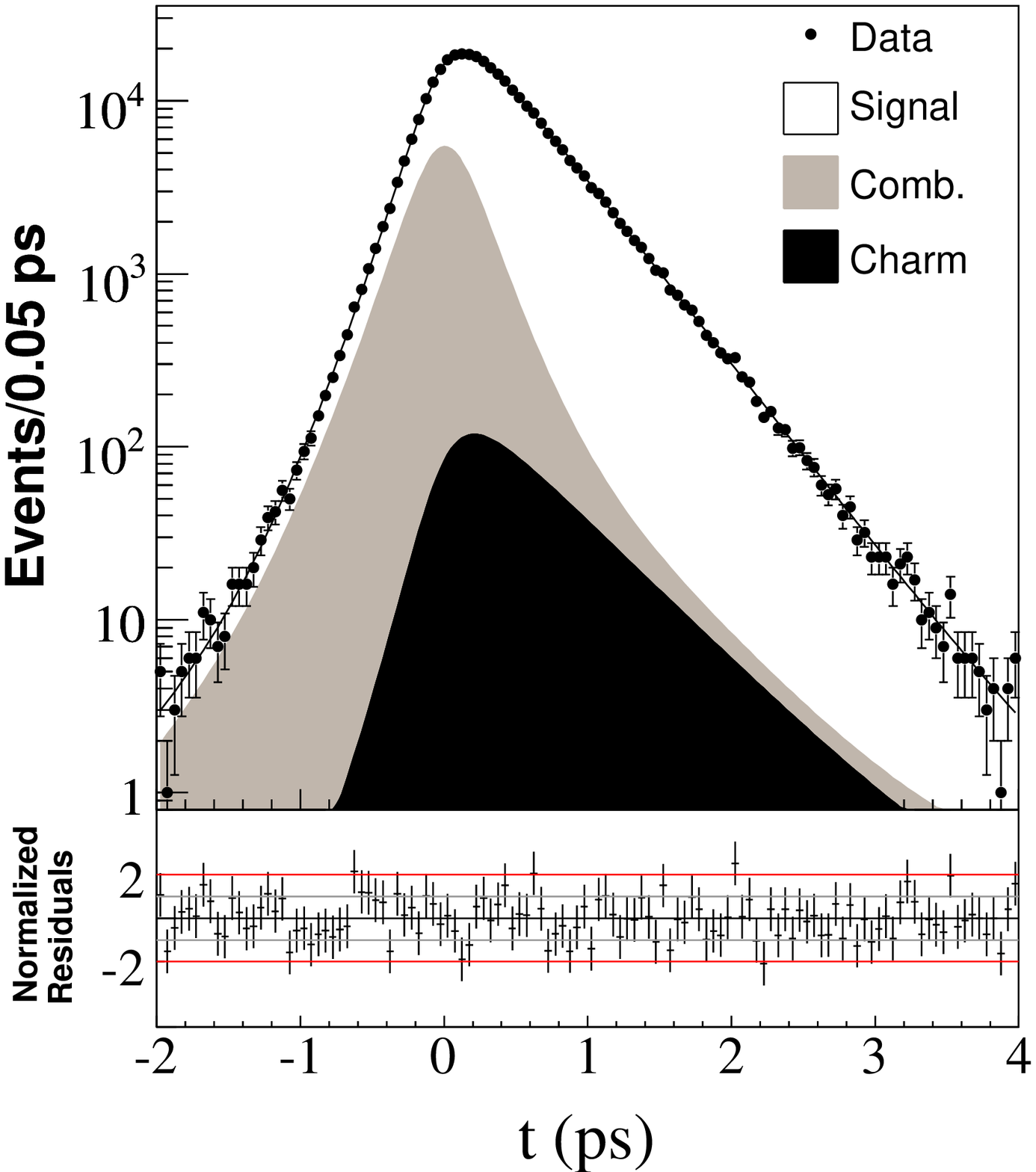} \\
\end{tabular}
\caption{\label{fig:KK-BaBar-UnTagged} BaBar $\Dz \to K^+ K^-$ untagged analysis~\cite{ref:KKpipi-BaBar-2009}.
(Left) $\Dz \to K^- \pi^+$ and (Right) $\Dz \to K^- K^+$ decay-time distributions with the data (points) and the projection of the lifetime fit overlaid.
The gray and black distributions represent the combinatorial and charm background contributions.
}
\end{center}
\end{figure}

Recently, BaBar has presented an untagged analysis using the same data sample of 384 fb$^{-1}$~\cite{ref:KKpipi-BaBar-2009}.
The proper-time distributions for the untagged $\Dz \to K^- \pi^+$ and $\Dz \to K^+ K^-$ samples are shown in 
Fig.~\ref{fig:KK-BaBar-UnTagged}, together with the projection of the simultaneous fit to these samples.
In this case, the samples contain $2710.2\times10^3$ and $263.6\times10^3$ signal events with purities of 94.2\% and 80.9\%,
for  $\Dz \to K^- \pi^+$ and $\Dz \to K^+ K^-$ respectively.
The measured lifetimes are 
$\langle \tau_{K^+K^-} \rangle = 405.85 \pm 1.0 {\rm (stat.)}$ fs 
and 
$\tau_{K^-\pi^+} = 410.39 \pm 0.38 {\rm (stat.)}$ fs,
yielding $y_{CP} = [1.12\pm0.26{\rm (stat.)}\pm0.22{\rm (syst.)}]\%$, which excludes no-mixing with $3.3\sigma$.
In this analysis, since the initial flavor of the decaying \Dz does not need to be identified, no $D^{*+}$ reconstruction 
is required, increasing significantly the reconstruction efficiency but increasing the amount of background. 
To minimize it, the lifetime fit is performed in a narrow \Dz mass region around the nominal \Dz mass. The proper-time distribution for the 
main background component (combinatorial) is estimated from sideband \Dz mass regions, while for the small admixture of 
misreconstructed charm decays it is obtained from the simulation.
Combining the tagged and untagged results taking into account both statistical and systematic uncertainties, BaBar finds
$y_{CP} = [1.16\pm0.22{\rm (stat.)}\pm0.18{\rm (syst.)}]\%$. Summing statistical and systematic uncertainties in quadrature,
the significance of this measurement is $4.1\sigma$.

Yet another lifetime-difference analysis uses the $K_S^0 K^+ K^-$ final state, where the $\phi K_S^0$ region and its sidebands 
are examined to extract CP-odd and CP-even amplitudes. This is in effect a measurement of the lifetime in the CP-even and CP-odd 
parts of the $K_S^0 K^+ K^-$ Dalitz plot. Using 673 fb$^{-1}$, Belle has 
measured $y_{CP} = [0.11\pm0.61{\rm (stat.)}\pm0.52{\rm (syst.)}]\%$~\cite{ref:KsKK-Belle}. This is done by measuring
the mean lifetime $\tau_{ON}$ in the $\phi K_S^0$ region (mainly CP-odd) and the mean lifetime $\tau_{OFF}$ in the sidebands (mainly CP-even), 
along with the corresponding fractions $f_{ON}$ and $f_{OFF}$ of CP-even events in these regions. The lifetime asymmetry in these regions can 
then be related to $y_{CP}$,
\begin{eqnarray}
\label{eq:BelleKsKK}
\frac{\tau_{OFF}-\tau_{ON}}{\tau_{OFF}+\tau_{ON}} = y_{CP} \frac{f_{ON}-f_{OFF}}{1+y_{CP}(1-f_{ON}-f_{OFF})},
\end{eqnarray}
from which relation the latter is then determined. 
The main systematic uncertainties in this analysis come from {\it ON-OFF} differences in
the proper-time resolution function and the selection criteria, while the uncertainty from the Dalitz model 
assumptions needed to evaluate the CP-even content is negligible ($0.01\%$).

\begin{figure}[htb!]
  \begin{minipage}{1.00\hsize}
   \begin{center}
     \includegraphics[width=.98\textwidth]{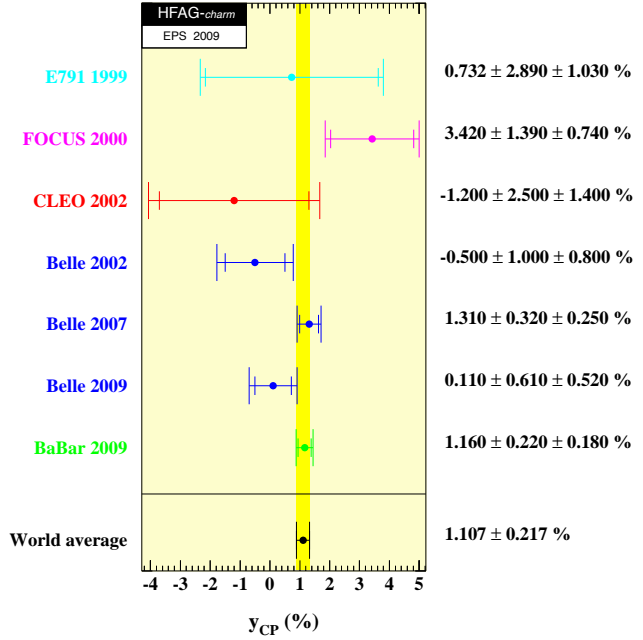}
      \caption{Summary of $y_{CP}$ measurements. The average $y_{CP}=(1.107\pm0.217)\%$ differs significantly from zero~\cite{ref:hfag}.}
    \label{fig:yCPsummary}
   \end{center}
  \end{minipage}
\end{figure}

Figure~\ref{fig:yCPsummary} summarizes all the available $y_{CP}$ results. The combination of all these measurements is performed
by the Heavy Flavor Averaging Group (HFAG)~\cite{ref:hfag}, and yields $y_{CP}=(1.107\pm0.217)\%$, 
which differs significantly (about $5\sigma$) from zero. 
The combined lifetime asymmetry is $A_\tau = (-0.123\pm0.248)\*$, thus there is no evidence for CPV.

\section{$\Dz \to K_S^0 \pi^+ \pi^-$ decays} 

The last employed technique to study \Dz-\Dzb mixing involves the multi-body final state $K_S^0 \pi^+ \pi^-$. As in the
case of the WS $\Dz \to K^+ \pi^- \pi^0$ analysis, the decay rate is a function of both
the Dalitz plot variables $s_+ = m^2_{K_S^0 \pi^+}$ and $s_- = m^2_{K_S^0 \pi^-}$, and the \Dz decay proper time.
With the usual approximations,
\begin{eqnarray}
\label{eq:Kspipi}
\frac{\Gamma(\Dz \to f)(s_+,s_-,t)}{e^{-\Gamma t} } = |A_f|^2 + ~~~~~~~~~~~~~~~~~~ \\
   |\overline{A}_f|^2 (\Gamma t)^2 \frac{R_M + |A_f|^2(y^2-x^2)/2}{2} + (\Gamma t) |A_f| |\overline{A}_f| y', \nonumber
\end{eqnarray}
with $y' = y \cos [ \delta_f(s_+,s_-)+\phi] - x \sin [\delta_f(s_+,s_-) + \phi]$, 
where $\delta_f(s_+,s_-) = {\rm arg} [ A_f^*(s_+,s_-) \overline{A}_f(s_+,s_-) ]$ is the relative strong phase between 
the \Dz and \Dzb decay amplitudes to the same final state $f=K_S^0 \pi^+ \pi^-$. 
Here, and contrary to the WS $\Dz \to K^+ \pi^- \pi^0$ case,
the strong phase $\delta_f$ is fixed 
by the fact that the \Dz and \Dzb Dalitz plots are identical 
(the $s_+$ and $s_-$ axes are just interchanged), assuming CP is not violated in the \Dz decay. Thus the analysis is free of 
unknown phases, providing an unique method to simultaneously measure the interfering \Dz and \Dzb amplitudes, 
the mixing parameters $x$ and $y$ (without rotations, and also their signs), and even the CP-violating parameters $\phi$ and $|q/p|$. 

\begin{figure}[htb!]
\begin{center}
  \includegraphics[width=0.4\textwidth]{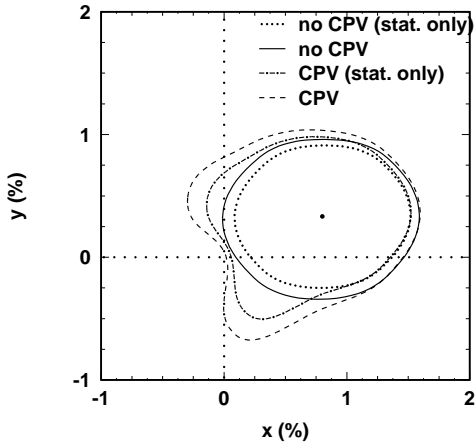}
\vskip-1.0cm
\caption{\label{fig:Kspipi-Belle-contours} Belle $\Dz \to K_S^0 \pi^+ \pi^-$ analysis~\cite{ref:Kspipi-Belle}.
95\% CL contours for $(x,y)$: dotted (solid) corresponds to statistical (statistical and systematic) contour for no CPV,
and dash-dotted (dashed) corresponds to statistical (statistical and systematic) contours for the CPV-allowed case.
The point is the best fit result for the no CPV case.
}
\end{center}
\end{figure}

The CLEO experiment pioneered this analysis using only 9 fb$^{-1}$ of 
data~\cite{ref:Kspipi-CLEO}, obtaining
the constraints $(-4.5<x<9.3)\%$ and $(-6.4<y<3.6)\%$ at 95\% CL. Using 60 times more data, Belle has also performed this analysis~\cite{ref:Kspipi-Belle}, 
first assuming CP conservation and subsequently allowing for CP violation. 
The amplitudes $A_f$ and $\overline{A}_f$ are described using 
a coherent sum of 18 amplitudes, 
dominated by CF $\Dz \to K^{*-} \pi^+$, DCS $\Dz \to K^{*+} \pi^-$ and CP $\Dz \to K_S^0 \rho^0$ decays.
Assuming negligible CP violation, 
$x = [0.80\pm0.29{\rm (stat.)}^{+0.09}_{-0.07}{\rm (syst.)}^{+0.10}_{-0.14}{\rm (model)}]\%$
and
$y = [0.33\pm0.24{\rm (stat.)}^{+0.08}_{-0.12}{\rm (syst.)}^{+0.06}_{-0.08}{\rm (model)}]\%$. This 
corresponds to a significance of $2.2\sigma$ from the no-mixing hypothesis. 
Figure~\ref{fig:Kspipi-Belle-contours} shows
both the statistical-only and overall contours for both the CPV-allowed and the CP-conservation cases. 
No evidence for CP violation is found.


\section{Wrong sign semileptonic decays} 

The most straightforward although not the most sensitive way to search for charm mixing is to use WS semileptonic decays, 
for instance $\Dz \to K^{(*)+} \ell^- \bar{\nu}_l$~\cite{ref:semilep-E791-CLEO,ref:semilep-BaBar,ref:semilep-Belle}.
In this case WS combinations 
can only occur through mixing, 
\begin{eqnarray}
\label{eq:WSsemileptonic}
\Gamma(\Dz \to f)(t) \propto e^{-\Gamma t} (\Gamma t)^2 R_M.
\end{eqnarray}
Therefore, these final states are only sensitive to 
$R_M \sim {\cal O}(10^{-4})$.
Using semileptonic decays for mixing searches involves the measurement of the time-dependent or time-integrated
rate for the WS decays. 
The main experimental challenge in these analyses is the limited mass resolution on $\Delta m$ due to the presence of neutrinos.
Significant improvements on $\Delta m$ resolution are obtained applying kinematic constraints on the invariant mass of the
neutrino and the kaon-lepton-neutrino system.
The best current limits are from BaBar, $R_M < 0.12\%$~\cite{ref:semilep-BaBar}, 
and Belle, $R_M<0.06\%$~\cite{ref:semilep-Belle}, both at 90\% CL, 
using 344 fb$^{-1}$ and 492 fb$^{-1}$ of data, respectively.

\section{Combined results} 

\begin{figure}[htb!]
  \begin{minipage}{1.00\hsize}
   \begin{center}
     \includegraphics[width=.80\textwidth]{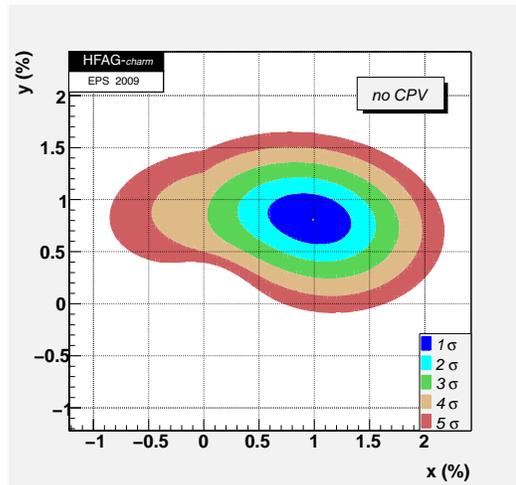}
      \caption{Two-dimensional contours for the mixing parameters $(x,y)$ from the HFAG combination~\cite{ref:hfag} using all available measurements, 
for the case of CP conservation. The no-mixing point is excluded with a significance clearly exceeding $5\sigma$.}
    \label{fig:xyn2d}
   \end{center}
  \end{minipage}
\end{figure}

The task of combining the wide variety of charm mixing results
is done by the HFAG~\cite{ref:hfag}. Figure~\ref{fig:xyn2d} shows the $(x,y)$ contours of the collective experimental
data, for the case of CP conservation. The central values $x=(0.989\pm0.241)\%$ and $y=(0.809\pm0.160)\%$ exclude the no-mixing point with
$10.2\sigma$. Other relevant combined parameters are $R_D = (0.3360\pm0.0084)\%$, $\delta_{K\pi}=0.44\pm0.17$~rad, and
$\delta_{K\pi\pi^0}=0.24\pm0.37$~rad. When CP violation is allowed, the mixing parameters remain basically unchanged,
$x=(0.976\pm0.249)\%$, $y=(0.833\pm0.160)\%$, $R_D = (0.3367\pm0.0086)\%$, $\delta_{K\pi}=0.46\pm0.17$~rad, and
$\delta_{K\pi\pi^0}=0.26\pm0.37$~rad, and the following values for the CP-violating parameters are obtained: 
$|q/p|=0.866\pm0.160$, $\phi = -0.148 \pm 0.126$~rad, and $A_D = (-2.2\pm2.4)\%$.

\section{Summary and conclusions} 

More than thirty years after the discovery of the \Dz meson~\cite{ref:goldhaber} and the first theoretical discussion
on mixing and CPV in the charm sector~\cite{ref:Pais1975}, BaBar, Belle and CDF Collaborations have provided
compelling experimental evidence for \Dz-\Dzb mixing. Collective experimental data favor the mixing hypothesis 
at $10.2\sigma$ level (including systematic uncertainties). The mixing measurement from \Dz lifetime differences 
$y_{CP}$ is significantly positive (about $5\sigma$), indicating that the $|D_1\rangle$ eigenstate ($\approx$ CP-even)
has a shorter lifetime than the $|D_2\rangle$ eigenstate ($\approx$ CP-odd).

However, no observation (more than $5\sigma$) in a single measurement has yet been presented. 
In addition, to date there is only one direct measurement (from $\Dz \to K_S^0 \pi^+ \pi^-$ decays) of $x$ and $y$ free of rotations (thus the only indication of
the $x$ and $y$ relative sign), which do not differ significantly from zero but do affect the combination of all mixing 
measurements due to the large uncertainties that arise from the unknown phases $\delta_f$ that are 
inherent in other determinations. This is especially true for $x$.
Clearly, more such measurements are a high priority and are foreseen in the future.

The measured values of the mixing parameters $x \approx y \approx 1\%$ are about compatible with SM expectations, 
although with large theoretical uncertainties. There is no evidence for CP violation in \Dz mesons, either in mixing, 
in decay or in interference.

Significant improvements in precision are foreseen in the short term with the analysis of complete data sets from current 
facilities (B factories and Tevatron). In the long term, facilities about to start or just starting (LHCb and BESIII) have the potential
to improve the precision on the mixing parameters in about a factor 5 and look deeper into CP violation searches.
In the longer term, SuperB factories could be the last opportunity to observe CP violation in \Dz-\Dzb mixing~\cite{ref:SuperB}
and search for NP in FCNC in the down-quark sector.

\section{Acknowledgements} 

I would like to thank the organizers of PIC2009 for the invitation to give this review and for having set the atmosphere for such interesting meeting.

I am grateful to my colleagues of the BaBar Collaborartion who helped preparing this talk and proceedings, with special thanks to Nicola Neri.




\begin{thebibliography}{99}
\bibitem{ref:phase-convention} The phase convention is chosen so that $CP |\Dz\rangle = |\Dzb\rangle$ and $CP |\Dzb\rangle = |\Dz\rangle$.
\bibitem{ref:chargeconj} Reference to the charge-conjugate state is implied here and throughout the text unless otherwise stated.
\end{thebibliography}
\end{document}